\documentclass[]{spie}  

\usepackage{amsmath,amsfonts,amssymb}
\usepackage{graphicx}
\usepackage[colorlinks=true, allcolors=blue]{hyperref}
\usepackage{float}
\usepackage{tabularx}
\usepackage{booktabs}
\usepackage[labelfont={color=black,bf}]{caption}
\usepackage{subcaption}
\usepackage[dvipsnames]{xcolor}
\definecolor{mybeige}{RGB}{169, 149, 123}
\definecolor{myyellow}{RGB}{240, 200, 0}
\definecolor{mycyan}{RGB}{0, 170, 170}
\definecolor{mygreen}{RGB}{0, 220, 0}
\usepackage{hyperref}

\title{Convolutions, Transformers, and their Ensembles for the Segmentation of Organs at Risk in Radiation Treatment of Cervical Cancer}

\author[a]{Vangelis Kostoulas}
\author[b]{Peter A.N. Bosman}
\author[a]{Tanja Alderliesten}
\affil[a]{Dept. of Radiation Oncology, Leiden University Medical Center, P.O. BOX. 9600, 2300 RC Leiden, the Netherlands}
\affil[b]{Evolutionary Intelligence Group, Centrum Wiskunde \& Informatica, P.O. BOX. 94079 1090 GB Amsterdam, the Netherlands}

\authorinfo{Further author information: (Send correspondence to Vangelis Kostoulas or Tanja Alderliesten)\\Vangelis Kostoulas: E-mail: E.Kostoulas@lumc.nl,\\  Tanja Alderliesten: E-mail: T.Alderliesten@lumc.nl}

\graphicspath{ {./images/} }

\begin{document} 
\maketitle

\begin{abstract}
Segmentation of regions of interest in images of patients, is a crucial step in many medical procedures. Deep neural networks have proven to be particularly adept at this task. However, a key question is what type of deep neural network to choose, and whether making a certain choice makes a difference. In this work, we will answer this question for the task of segmentation of the Organs At Risk (OARs) in radiation treatment of cervical cancer (i.e., bladder, bowel, rectum, sigmoid) in Magnetic Resonance Imaging (MRI) scans. We compare several state-of-the-art models belonging to different architecture categories, as well as a few new models that combine aspects of several state-of-the-art models, to see if the results one gets are markedly different. We visualize model predictions, create all possible ensembles of models by averaging their output probabilities, and calculate the Dice Coefficient between predictions of models, in order to understand the differences between them and the potential of possible combinations. The results show that small improvements in metrics can be achieved by advancing and merging architectures, but the predictions of the models are quite similar (most models achieve on average more than 0.8 Dice Coefficient when compared to the outputs of other models). However, the results from the ensemble experiments indicate that the best results are obtained when the best performing models from every category of the architectures are combined.
\end{abstract}


\section{Introduction}
\label{sec:intro}

\begin{figure}
\centering
\includegraphics[width=11.0cm,height=5cm]{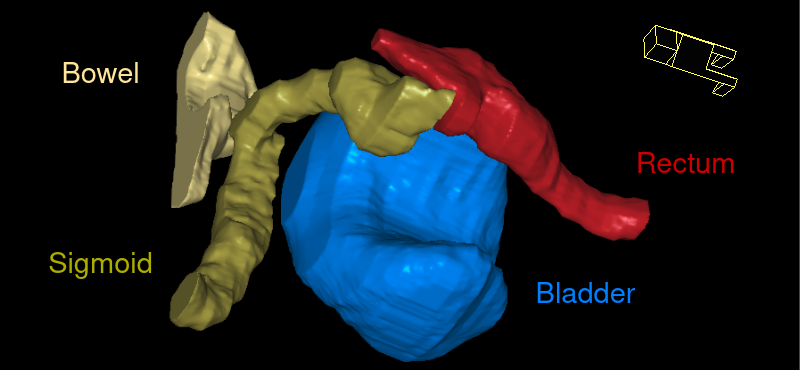}
\vspace{0.5cm}
\caption{3D visualization of the OARs in brachytherapy of cervical cancer. At the top right corner of the figure, the orientation of the patient is shown.}
\label{fig:oars}
\end{figure}

Radiation treatment requires as initial steps the acquisition of a patient image (e.g., Magnetic Resonance Imaging (MRI), Computed Tomography (CT) scan), and then the delineation of regions of interest (e.g., organs, tumour) based on the given image. For example, in internal radiation treatment planning of cervical cancer (i.e., brachytherapy), the delineation of the bladder, bowel, rectum, and sigmoid is essential, since these are the Organs At Risk (OARs) (\autoref{fig:oars}). These delineations are used for radiation dose distribution optimization. The goal of such optimization is to minimize the amount of radiation dose delivered to the OARs while maximizing tumour coverage.

Performing delineations manually requires a considerable amount of time and effort and is biased by observer variation \cite{obsvar1, obsvar2, obsvar3, obsvar4}. Therefore, there is a need to perform these tasks faster and more robustly. In the last years, deep learning techniques have shown to be particularly effective in medical image segmentation tasks \cite{unet, vnet, nnunet, swinunetr2}. Convolutional Neural Networks (CNNs) were the undisputed state-of-the-art tools for vision tasks, until the arrival of efficient vision Transformer-based networks, which achieve a very competitive performance \cite{vit, swintransformer}. CNNs excel at capturing local relationships, while Transformers can capture global relationships. For this reason, researchers have also proposed architectures which utilize both convolutions from CNNs and self-attention mechanisms from Transformers, in order to extract both local and global features, respectively \cite{coatnet, cvt}.

For a medical image segmentation task, the standard approach is to first select an existing network's architecture, known to work well for segmentation in general, and then to manually tune it for a new task. It is, however, unknown a priori what architecture will work best. In this work, we experiment with state-of-the-art architectures from different architecture categories, and gain insight into which architectures work well for the segmentation of the OARs in cervical cancer patients treated with brachytherapy. We also experiment with new architectures to investigate possible performance improvements by advancing and combining architectures.

Additionally, after training networks on a target dataset, it is common among practitioners to combine network's outputs (ensembling), as it increases performance by reducing variation \cite{ensemble}. The most common approach in this case is to take the best performing network and train it multiple times, to create multiple models which will then be combined to form the final ensemble. We experiment with the ensembling step, and next to using an ensemble from versions of the best performing model, we also create ensembles from different state-of-the-art architectures. 

To summarize, in this work we shed light on how state-of-the-art, and also new combinations of, architectures perform on a specific segmentation task, how they differ, and what is the potential of using an ensemble of different architectures.

\section{Related Work}
\label{sec:relatedwork}

Undoubtedly, one of the pioneering architectures for medical image segmentation is the U-Net \cite{unet} model, which achieves impressing results and can be applied to a vast range of different segmentation tasks. This model consists of a convolutional encoder which encodes an image to a set of feature maps, and a convolutional decoder which receives the image encoding and transforms it into a segmentation mask. Additionally, the layers of the encoder are passed to the decoder layers through skip connections, further enhancing the performance of the model by providing information to the decoder that is lost from the downsampling. After the U-Net, the V-Net \cite{vnet} was proposed to solve 3D medical image segmentation tasks. The model has an architecture similar to U-Net, but uses 3D convolutions instead of 2D, and generally can achieve better performance on 3D tasks than 2D approaches, which are based on dividing the 3D volumes into 2D slices. By processing directly the 3D images, the memory requirements are increased significantly, but it does allow to also capture relationships between all the dimensions of a 3D volume, hence the performance improvement of 3D models. Finally, one of the state-of-the-art works in the field of medical segmentation right now is the nnUNet \cite{nnunet}, in which both 2D and 3D U-Nets are combined to solve 3D medical tasks. Methods to automate the hyper parameter selection are also suggested \cite{nnunet}, resulting in a system that can adapt to new tasks and perform decently well, without having to optimize the hyper parameter set. This system achieved one of the top performances in several medical tasks at the time of publication, by using only simple U-Net architectures without other modifications, suggesting that the exact architecture of the networks does not impact that much the overall performance and instead the hyper parameter selection does.

After the U-Net was published, researchers tried to design more advanced architectures, with the purpose of achieving better performance. A natural addition to the original U-Net architecture is residual learning \cite{residuals}, which was used for example in the V-Net. Furthermore, the addition of extra convolutions in the skip connections could also enhance the performance of networks, by closing the distribution gap between encoder and decoder layers \cite{unetplus}. Finally, Atrous Spatial Pyramid Pooling (ASPP) \cite{deeplab} in the bottleneck layer is a commonly used technique. In the bottleneck layer, the field of view of a convolution kernel covers a large part of the image representation, while this layer does not have much influence on the required computation time. It might therefore be wise to use a combination of convolutions and/or dilated convolutions there.

Recently, however, another architecture has revolutionized the deep learning field, the Transformer architecture \cite{transformer}. Transformers were first proposed to tackle natural language processing tasks, but recently it was reported that Transformers can also process images if we divide the images into patches and consider those patches as words in a sentence \cite{vit}. Since then, a lot of Transformer-based architectures have been proposed to solve vision problems, challenging the long dominance of convolutional neural networks.

An architecture called the UNETR \cite{unetr}, has been recently proposed as the first architecture which uses the original Vision Transformer \cite{vit} as encoder for segmentation, and achieved interesting results. After the UNETR model, the Swin UNETR model \cite{swinunetr} was introduced, which contains a Swin Transformer \cite{swintransformer} as encoder. Additionally, it was shown that the Swin UNETR together with self-supervised pretraining \cite{swinunetr2}, achieves state-of-the-art results on a variety of medical datasets, including most of the datasets from the Medical Segmentation Decathlon challenge \cite{msd}. The UNETR model contains an original Vision Transformer, which is a powerful model. However, small patch sizes cannot be used due to computational complexity, which might be needed in tasks like segmentation, where we want pixel-level precision. The Swin UNETR is, in a way, simulating the behavior of a CNN, but instead of convolutions to extract features, it efficiently uses attention-based feature extraction. One downside of the Swin UNETR might be that it calculates the attention only inside windows of patches.

It seems though, that for dense prediction tasks like object detection and segmentation, convolutions work really well, leading to fast convergence and being able to learn from very small data sets. Additionally, most of the state-of-the-art networks for segmentation rely on a convolutional decoder, as more robust and general techniques have not yet been developed or at least proven to work as well. In the medical field, the most common scenario is having to work with very small datasets, and Transformers need typically a lot of data to be trained, as they lack some very useful inductive biases that are present in convolutions, like translation invariance. For these reasons, the combination of Transformers and convolutions has gained interest, in order to design networks that can capture both global and local relationships, and to utilize the best of both. Some well known models are the CoatNet \cite{coatnet}, which achieved state of the art results on the ImageNet benchmark, and the CvT \cite{cvt}. The combination of Transformers and convolutions is also explored in medical image segmentation, with works like \citenum{transunet, transfuse}.

Finally, we would like to mention that the delineation of the OARs during internal radiation treatment of cervical cancer patients with deep learning models, has been studied by several researchers \cite{cervix1, cervix2, cervix3, cervix4}. However, to our knowledge, only one work considers MRI images, while in the other studies CT scans have been used. Furthermore, there is no work where the bowel is an organ of interest, there is no work showing the performance of Transformer models on this task, and the ensembling of different architectures has not been studied yet.

\section{Networks}
\label{sec:networks}

\subsection{Convolution-Based Networks}
\label{sec:cnns}

\begin{figure}[H]
\centering
\includegraphics[width=14.0cm,height=6cm]{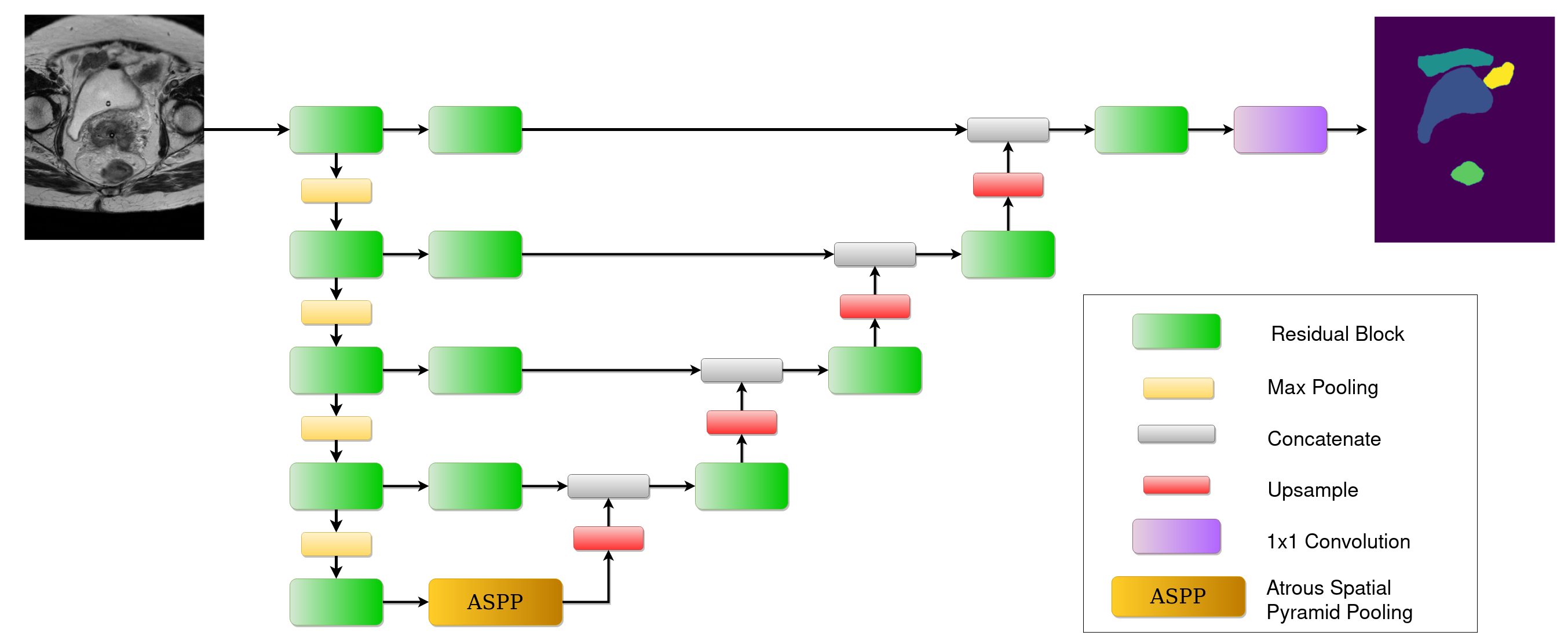}
\vspace{0.5cm}
\caption{CUnet architecture.}
\label{fig:cunet}
\end{figure}

In our work, we implement the U-Net architecture \cite{unet}, but we use GeLU instead of ReLU activations and bilinear interpolation for upsampling instead of transposed convolutions. We also design an advanced CNN, the Combined Unet (CUnet) (\autoref{fig:cunet}), which is based on the original U-Net, and also includes residuals, additional convolutions in the skip connections, and an ASPP block in the bottleneck. The output of the final encoder layer passes through 4 dilated convolutions (the ASPP block) with dilation rates 1, 2, 3, and 4, which are then concatenated and passed through a convolution with kernel size 1 to reduce the number of channels.

\subsection{Transformer-Based Networks}
\label{sec:transformers}

Inspired by the UNETR and the Swin UNETR, we propose a new Transformer-based segmentation architecture, the MultiScale UneTr (MSUneTr). \hspace{-0.03cm}The architecture considers multiple Vision Transformers as the encoder layers, with each one of them having a patch size equal to $2^n \times 2^n$, where $n \in \mathbb{N}$, is the corresponding layer index. The layers of the encoder are not connected and instead pass their output directly to the corresponding decoder layer. With this architecture, extra transposed convolutions are not needed in order to reshape the features of the encoder, as is the case in the UNETR. This is because different patch sizes are used instead of a fixed one. The MSUneTr is developed based on the idea to redesign the UNETR architecture in order to integrate the concept of multi-scale processing (processing of images in different resolutions). It is also worth mentioning that this architecture takes into account the relationships of all possible patches, in contrast with the Swin UNETR, where the attention is calculated only inside windows of patches.

\begin{figure}[H]
\centering
\includegraphics[width=12cm,height=4cm]{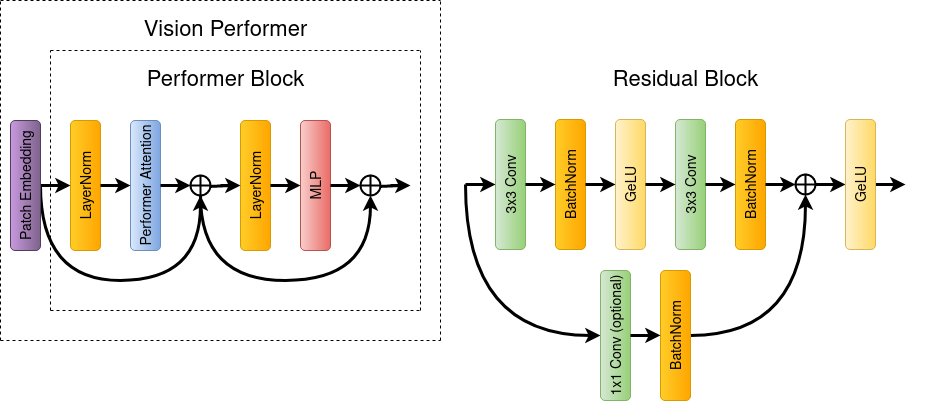}
\vspace{0.5cm}
\caption{Blocks used in the MSUneTr. Left: Performer block with patch embedding (together forming a Vision Performer). Right: main convolutional block used in the model.}
\label{fig:blocks}
\end{figure}

\begin{figure}[H]
\centering
\includegraphics[width=15.0cm,height=6cm]{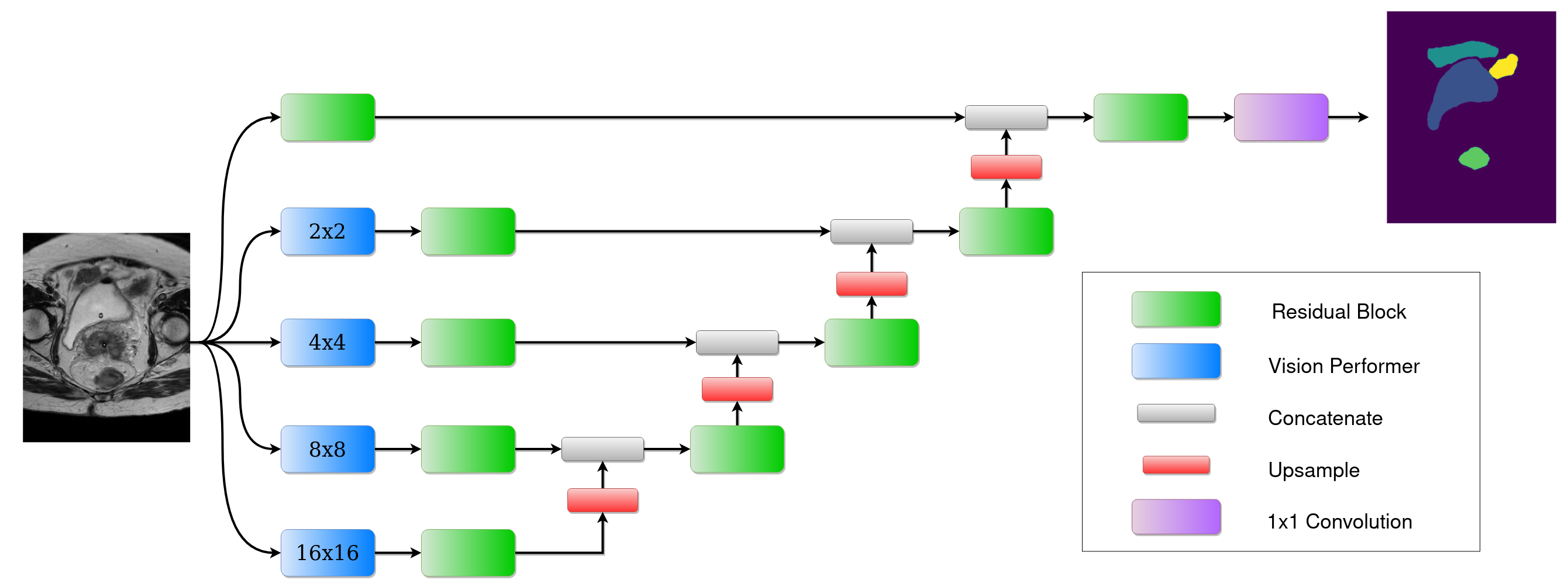}
\vspace{0.5cm}
\caption{MSUneTr architecture.}
\label{fig:msunetr}
\end{figure}

The smaller the patch size in Vision Transformers, the larger the input sequence, the more computationally expensive the attention calculation (having a quadratic complexity with respect to the input), resulting in memory inefficiency even for very small image sizes. \hspace{-0.05cm}Therefore, we propose to replace regular Transformers with Performers \cite{performer}, which estimate the attention kernels, and create a network which is much faster, requires far less memory, and also maintains the same performance. In theory, every Vision Performer (Performer with a patch embedding layer) consists of an arbitrary number of layers, but here we use just 1 Performer block, to demonstrate that this idea works. In \autoref{fig:blocks} the main blocks used in the MSUneTr are illustrated, and the whole architecture is shown in \autoref{fig:msunetr}.

\subsection{Mixed Networks}
\label{sec:mixednetworks}

\begin{figure}[H]
\centering
\includegraphics[width=17cm,height=7cm]{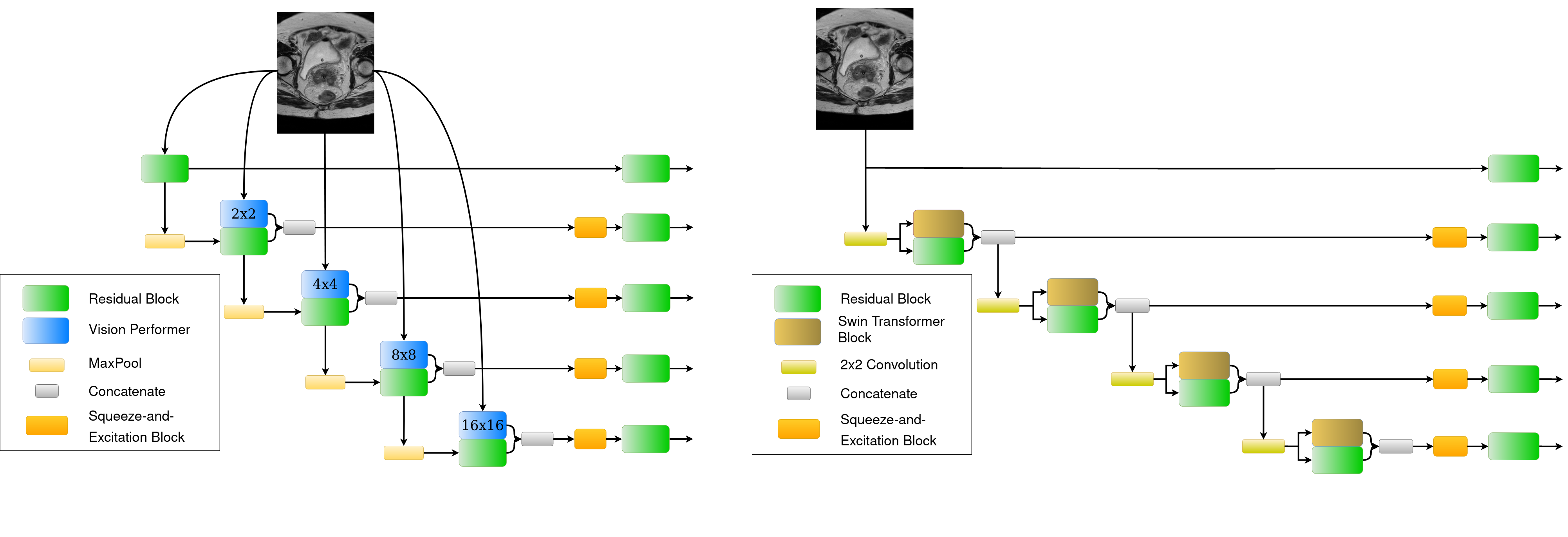}
\vspace{-15pt}
\caption{DeceptiConv encoder (left), and SwinConvNet encoder (right).}
\label{fig:mixednets}
\end{figure}

We consider two simple ways of combining convolutions and Transformers (\autoref{fig:mixednets}). We merge the MSUneTr with a similar network to CUnet, through the concatenation of the Vision Performer outputs with the respective outputs of the convolutional encoder. We call this network DeceptiConv. We also merge the Swin UNETR with a similar network to CUnet, by concatenating the Swin Transformer \cite{swintransformer} blocks with the corresponding outputs of the convolutional encoder. We call this network SwinConvNet. We use $2\times2$ convolutions to learn the downsampling operations in SwinConvNet, instead of having a fixed operation, in order to create a more generic approach with the potential to fit better to specific training sets. To let the networks decide which features are the most useful, we also add Squeeze-and-Excitation blocks \cite{se} after the concatenation. Both networks (and also the CUnet, and the MSUneTr) use exactly the same decoder.

Notice that, in both networks, we can use either Performer blocks, or Swin Transformer blocks. The main difference is that, in the DeceptiConv we use patch embedding layers with different patch sizes in order to pass the input image into the Performer blocks, while in the SwinConvNet, the role of the patch embedding layers is assigned to every downsampling convolution of the encoder. The SwinConvNet is a more conventional way of combining Transformers with convolutions, and other researchers have proposed similar architectures (\citenum{transconver}).

\subsection{Summary of Networks}
\label{sec:details}

Bellow we summarize the main details associated with the various investigated networks in this work. The number of model parameters are provided in \autoref{table:table1}.

\begin{itemize}

\item \textbf{U-Net}: based on the original architecture, but with GeLU activations instead of ReLU and bilinear interpolation for upsampling instead of transposed convolutions. We use multiples of 48 as the number of channels of every convolution operation. The bottleneck layer consists of 2 convolutions, the first producing 768 channels, and the second 384 channels.

\noindent
\item \textbf{CUnet}: advanced convolutional architecture based on the U-Net, with residual convolutional blocks, additional residual blocks in the skip connections, and an ASPP block in the bottleneck. We use multiples of 48 as the number of channels of every convolution operation. The output of the layer before the bottleneck passes through 4 dilated convolutions with dilation rates 1, 2, 3, and 4, each one producing 384 channels. Then all the feature maps are concatenated and passed through a final convolution with kernel size 1 to reduce the number of channels.

\noindent
\item \textbf{UNETR}: model taken from the MONAI library version 0.8.1 with the default parameters.

\noindent
\item \textbf{Swin UNETR}: model taken from the MONAI library version 0.8.1 with feature size 48, and the rest of the parameters are the same as the default.

\noindent
\item \textbf{MSUneTr}: multiple Vision Performers with different patch sizes as the encoder layers. Each Vision Performer has only one layer, that produces multiples of 48 feature maps according to the layer index. Additional convolutions are used in the skip connections.

\noindent
\item \textbf{DeceptiConv}: combined MSUneTr with CUnet (without the ASPP block). We use multiples of 48 as the number of feature maps produced from every layer. Squeeze-and-Excitation blocks are used before the additional convolutions in the skip connections.

\noindent
\item \textbf{SwinConvNet}: combined Swin UNETR with CUnet (without the ASPP block), with $2\times 2$ convolutions for downsampling. We use multiples of 48 as the number of feature maps produced from every layer. Squeeze-and-Excitation blocks are used before the additional convolutions in the skip connections.

\end{itemize}

\begin{table}[H]
\captionsetup{font=footnotesize}
\centering
\caption{Number of parameters of the investigated models.}

\resizebox{\textwidth}{!}{%
\begin{tabular}{*{1}{>{\bfseries}l} *{7}{|>{\bfseries}c} }
\specialrule{.2em}{.1em}{.1em}
 Model & U-Net & CUnet & UNETR & Swin UNETR & MSUneTr & DeceptiConv & SwinConvNet \\
 \specialrule{.1em}{.05em}{.05em}
  Parameters & $10,188,773$ & $14,605,301$ & $87,118,837$ & $25,122,917$ & $7,979,765$ & $27,782,549$ & $27,106,037$ \\
\specialrule{.2em}{.1em}{.1em} 
\end{tabular}
}
\label{table:table1}
\end{table}

\section{Experiments}
\label{sec:experiments}

In this study, we assume that we do not have datasets available for pretraining, which is known to boost the performance of networks, and we can only train networks from scratch on the target datasets. Also, due to a limited amount of resources, we conduct experiments with 2D networks using only one GPU per experiment. For all the experiments, 1 NVIDIA V100 GPU and 1 NVIDIA A100 GPU are used, separately. Finally, to further validate our findings, we also conducted experiments on the Brain Tumour Dataset of the Medical Segmentation Decathlon \cite{msd}, a publicly available dataset (see \hyperref[sec:appendix]{appendix}). 

\subsection{Cervical Cancer Dataset}
\label{sec:cervix}
We retrospectively conduct experiments on a clinical dataset from the Leiden University Medical Center, a private dataset which contains T2-weighted MRI scans obtained with Philips Ingenia, Intera, and Achieva 1.5T or 3T MRI systems (Philips Healthcare, Best, The Netherlands) of patients with cervical cancer treated between 2008 and 2021. With each scan annotations are associated, which have been created as part of standard clinical care for the purpose of brachytherapy treatment planning. We removed patient images with artifacts or missing annotations, resulting in 194 patient scans in total, consisting of 7680 slices which were used as the training dataset. In this work, we focus on the following 4 OARs: the bladder, the rectum, the bowel, and the sigmoid (\autoref{fig:oars}).


At this point, we would like to mention that the accurate delineation of the rectum, sigmoid, and bowel is a very challenging problem for deep learning models. They are part of the intestines and are connected. The bowel represents the small and large intestines. The sigmoid is the terminal portion of the large intestine, connecting the descending colon with the rectum. \hspace{-0.1cm}These organs have very abstract shapes, and often because of poor contrast, it is hard to distinguish between these organ tissues and their surrounding tissues. Additionally, there are no global guidelines that specify exactly which part of the intestine corresponds to each sub-organ, which increases the observer variation for the boundaries defining the transition between the aforementioned organs.

\begin{figure}[H]
\centering
\includegraphics[width=17cm,height=6cm]{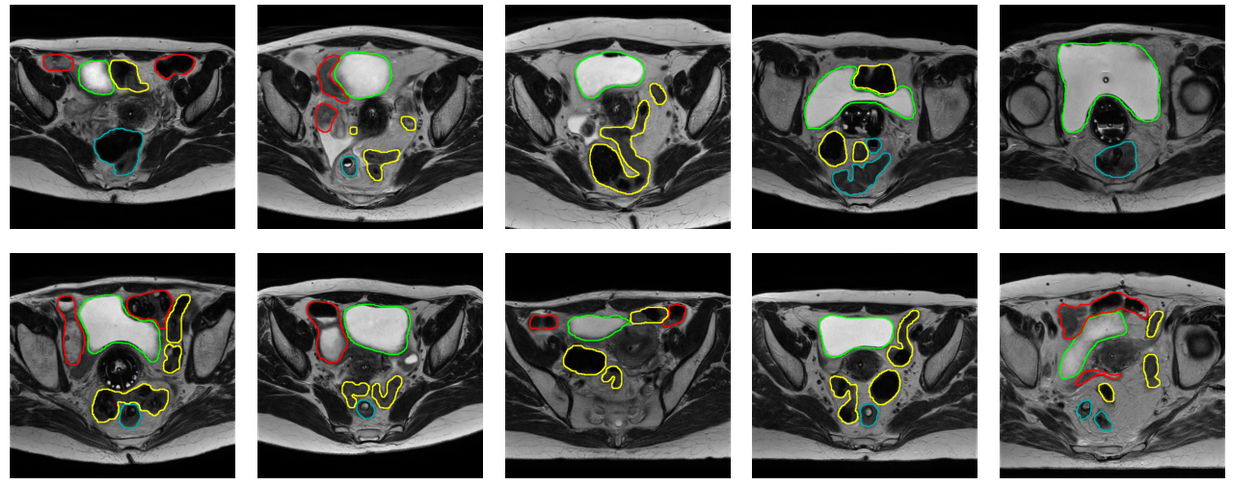}
\vspace{0.5cm}
\caption{\footnotesize{Slices of images from the Cervical Cancer Dataset. {\color{mygreen} \textbf{Green}}: bladder, {\color{Red} \textbf{Red}}: bowel, {\color{mycyan} \textbf{Cyan}}: rectum, {\color{myyellow} \textbf{Yellow}}: sigmoid.}}
\label{fig:examples}
\end{figure}

\subsection{Training and Inference}
\label{sec:training}

Patches of size $320 \times 320$ are cropped to train the networks, and to get the predictions of a model on a single image slice, we follow a sliding window approach with an overlap of 0.5. We adopted the preprocessing pipeline from \citenum{nnunet}, so all the data is cropped to the region of nonzero values, all patients are resampled to the median voxel spacing of their respective dataset, where third order spline interpolation is used for image data and nearest neighbor interpolation for the corresponding segmentation mask, and z-score normalization is applied to every patient image individually. Preliminary experiments indicated that transformations other than simple spatial transformations (specifically rotations and scaling) did not bring substantial improvements in terms of quantitative metrics, so we used only these simple transformations. We also have to mention that flips were not used for the Cervical Cancer Dataset, as the performance was decreased considerably, probably because the location of the objects in this task plays an important role.

In all the experiments, we use the AdamW optimizer with a learning rate of $3\cdot10^{-4}$ and a weight decay of $0.05$. We train all the models with a batch size of $16$ and a combination of Dice and cross-entropy loss. While training, we decrease the learning rate by a factor of $0.5$ if the training loss did not decrease for the last 3 consecutive epochs, until it is not smaller than $1\cdot10^{-5}$. 

In order to estimate the generalization performance of the networks, 5-fold cross validation experiments are performed with 100 epochs training per fold. Furthermore, we create ensembles of all possible combinations of models by averaging their output probabilities, to find out how the models synergize. Additionally, we create an ensemble from 5 trainings of the best performing model (in this case, the DeceptiConv), to compare it with the ensembles of multiple different architectures.

\subsection{Evaluation}
\label{sec:evaluation}

We evaluate all models by calculating the Dice Coefficient and the $95^{th}$ percentile of Hausdorff Distance between the predictions and the references. In order to gain more insight into the similarity of the investigated models, we also pairwise compute the Dice Coefficients between their predictions. 

For every model, in every k-fold round, we calculate metrics per class (i.e., OAR) for every patient. Then we average the results for all patients across classes. We collect the average values for every fold, and then we average them again across classes to get the final results. Additionally, we provide results from the best ensembles of models. For every patient, we receive the output probabilities of the models in the ensemble, and then we average them with equal weight. The final results for the ensembles are calculated by following the same process as for the single-model case.

To investigate the differences between the models and the ensembles in terms of statistical significance, we conduct two Wilcoxon paired tests, one for the best performing ensemble and the best performing model, and a second one for the two best performing models (performance according to quantitative metrics). Results with $p$-value $< 0.05$ were considered to be significant.

\section{Results}
\label{sec:results}

In \autoref{table:table2} the results from the 5-fold experiments of every single model, but also from the ensemble experiments are presented. The combination that achieves the best performance according to the average metrics (i.e., averaged over all OARs as presented in the Avg column in \autoref{table:table2}) is referred to as Ensemble 1 (DeceptiConv, SwinConvNet, CUnet, Swin UNETR, and U-Net), followed by Ensemble 2 (DeceptiConv, SwinConvNet, CUnet, Swin UNETR, U-Net, and UNETR), Ensemble 3 (DeceptiConv, SwinConvNet, CUnet, Swin UNETR, U-Net, and MSUneTr), and finally Ensemble All (all 7 models). The ensemble from the 5 DeceptiConvs is referred to as Ensemble DC.

\begin{table}[H]
\captionsetup{font=footnotesize}
\centering
\caption{Average Dice Coefficient $\pm$ standard deviation and $95^{th}$ percentile of Hausdorff Distance (HD) in mm $\pm$ standard deviation, on the validation sets, from models trained with 5-fold training, and the best performing ensembles on the Cervical Cancer Dataset.}

\resizebox{\textwidth}{!}{%
\begin{tabular}{*{1}{>{\bfseries}l} *{10}{|>{\bfseries}c} }
 \specialrule{.2em}{.1em}{.1em} 
 \multicolumn{1}{c|}{Model} & \multicolumn{2}{c|}{bladder} & \multicolumn{2}{c|}{bowel} 
 & \multicolumn{2}{c|}{rectum} & \multicolumn{2}{c|}{sigmoid} & \multicolumn{2}{c}{Avg} \\
 \specialrule{.1em}{.05em}{.05em}
  & Dice & HD & Dice & HD & Dice & HD & Dice & HD & Dice & HD \\
 \specialrule{.1em}{.05em}{.05em}
 U-Net & $0.9525$ & $5.7338$ & $0.6121$ & $24.4093$ & $0.8131$ & $14.7873$ & $0.7075$ & $25.2454$ & $0.7713$ & $17.5440$  \\
  & $\pm 0.0044$ & $\pm 0.8563$ & $\pm 0.0166$ & $\pm 0.9268$ & $\pm 0.0088$ & $\pm 1.6146$ & $\pm 0.0092$ & $\pm 1.5363$ & $\pm 0.0045$ & $\pm 0.8047$  \\
 CUnet & $0.9515$ & $5.3130$& $0.6316$ & $23.8194$ & $0.8198$ & $14.9128$ & $0.7108$ & $25.2929$ & $0.7784$ & $17.3345$  \\
  & $\pm 0.0049$ & $\pm 0.6390$ & $\pm 0.0242$ & $\pm 1.1534$ & $\pm 0.0131$ & $\pm 1.7234$ & $\pm 0.0069$ & $\pm 1.0201$ & $\pm 0.0093$ & $\pm 0.4818$  \\
 UNETR & $0.9439$ & $9.3933$ & $0.5650$ & $28.2959$ & $0.7853$ & $18.3888$ & $0.6457$ & $31.1964$ & $0.7350$ & $21.8186$  \\
  & $\pm 0.0097$ & $\pm 3.0756$ & $\pm 0.0366$ & $\pm 3.0608$ & $\pm 0.0140$ & $\pm 1.1880$ & $\pm 0.0186$ & $\pm 1.8578$ & $\pm 0.0170$ & $\pm 1.5400$  \\
 Swin UNETR & $0.9491$ & $3.3010$ & $0.6238$ & $32.2772$ & $0.8087$ & $10.6996$ & $0.6939$ & $32.9731$ & $0.7689$ & $19.8128$  \\
  & $\pm 0.0065$ & $\pm 1.8891$ & $\pm 0.0223$ & $\pm 3.6262$ & $\pm 0.0140$ & $\pm 1.2674$ & $\pm 0.0139$ & $\pm 2.9408$ & $\pm 0.0106$ & $\pm 1.6556$  \\
 MSUneTr & $0.9500$ & $8.7216$ & $0.5927$ & $27.0443$ & $0.8070$ & $16.9178$ & $0.6831$ & $29.6154$ & $0.7582$ & $20.5748$ \\
  & $\pm 0.0060$ & $\pm 3.1051$ & $\pm 0.0210$ & $\pm 1.7122$ & $\pm 0.0118$ & $\pm 1.5286$ & $\pm 0.0106$ & $\pm 3.9122$ & $\pm 0.0056$ & $\pm 1.6353$  \\
 DeceptiConv & $0.9533$ & $4.9334$ & $0.6402$ & $23.3072$ & $0.8156$ & $14.6143$ & $0.7157$ & $24.9631$ & $0.7812$ & $16.9545$ \\
  & $\pm 0.0040$ & $\pm 0.2559$ & $\pm 0.0236$ & $\pm 1.1639$ & $\pm 0.0098$ & $\pm 1.2970$ & $\pm 0.0061$ & $\pm 1.3864$ & $\pm 0.0071$ & $\pm 0.7422$  \\
 SwinConvNet & $0.9514$ & $5.5482$ & $0.6370$ & $24.6621$ & $0.8172$ & $14.4843$ & $0.7147$ & $24.9706$ & $0.7801$ & $17.4163$ \\
  & $\pm 0.0042$ & $\pm 1.2329$ & $\pm 0.0239$ & $\pm 0.9677$ & $\pm 0.0149$ & $\pm 1.5494$ & $\pm 0.0108$ & $\pm 1.5134$ & $\pm 0.0079$ & $\pm 0.7408$  \\
  \specialrule{.1em}{.05em}{.05em}
 Ensemble DC & $0.9550$ & $3.7969$ & $0.6543$ & $22.6561$ & $0.8246$ & $14.1167$ & $0.7295$ & $22.9784$ & $0.7908$ & $15.8870$  \\  
  & $\pm 0.0062$ & $\pm 0.7544$ & $\pm 0.0198$ & $\pm 1.0503$ & $\pm 0.0116$ & $\pm 1.1378$ & $\pm 0.0103$ & $\pm 1.7025$ & $\pm 0.0093$ & $\pm 0.6931$  \\
  \specialrule{.1em}{.05em}{.05em}
 Ensemble 1 & $0.9560$ & $4.2718$ & $0.6567$ & $21.7427$ & $0.8290$ & $13.8169$ & $0.7349$ & $22.8120$ & $0.7942$ & $15.6609$ \\
  & $\pm 0.0050$ & $\pm 0.5927$ & $\pm 0.0240$ & $\pm 1.2138$ & $\pm 0.0114$ & $\pm 1.5778$ & $\pm 0.0089$ & $\pm 1.3324$ & $\pm 0.0090$ & $\pm 0.6582$  \\
 Ensemble 2 & $0.9560$ & $4.4756$ & $0.6562$ & $21.5814$ & $0.8295$ & $13.8305$ & $0.7344$ & $22.2224$ & $0.7941$ & $15.5275$ \\
  & $\pm 0.0051$ & $\pm 1.0489$ & $\pm 0.0236$ & $\pm 0.9176$ & $\pm 0.0113$ & $\pm 1.6189$ & $\pm 0.0098$ & $\pm 1.2677$ & $\pm 0.0093$ & $\pm 0.8077$  \\
 Ensemble 3 & $0.9562$ & $4.4427$ & $0.6550$ & $21.5355$ & $0.8298$ & $13.9176$ & $0.7349$ & $22.3883$ & $0.7940$ & $15.5710$  \\
  & $\pm 0.0051$ & $\pm 0.9609$ & $\pm 0.0236$ & $\pm 1.1167$ & $\pm 0.0113$ & $\pm 1.4476$ & $\pm 0.0078$ & $\pm 1.5951$ & $\pm 0.0086$ & $\pm 0.7979$  \\
 Ensemble All & $0.9560$ & $4.4244$ & $0.6546$ & $21.4820$ & $0.8299$ & $13.9058$ & $0.7340$ & $22.3658$ & $0.7936$ & $15.5445$ \\
  & $\pm 0.0052$ & $\pm 1.0665$ & $\pm 0.0241$ & $\pm 0.9982$ & $\pm 0.0113$ & $\pm 1.5191$ & $\pm 0.0084$ & $\pm 1.3771$ & $\pm 0.0092$ & $\pm 0.7500$  \\
\specialrule{.2em}{.1em}{.1em} 
\end{tabular}
}
\label{table:table2}
\end{table}

After training all the models based on the reference annotations, we compute the Dice Coefficient between the outputs of 2 models, for every possible pair of them. The results are calculated in the same way as in the normal 5-fold experiments, with the only difference being that we replace the references with the predictions of a model. We report the average results, across all classes (i.e., OARs), for every pair of models (\autoref{table:table3}). From the results, we observe that almost all models (except the UNETR) achieve a Dice score of more than 0.82, which indicates that the predictions of the models are very similar.

\begin{table}[H]
\captionsetup{font=footnotesize, width=\textwidth}
\centering
\caption{Average Dice Coefficient $\pm$ standard deviation on the validation sets, between predictions of models trained with 5-fold training on the Cervical Cancer Dataset.}

\resizebox{\textwidth}{!}{%
\begin{tabular}{*{1}{>{\bfseries}l} *{7}{|>{\bfseries}c} }
 \specialrule{.2em}{.1em}{.1em} 
 \multicolumn{1}{c|}{} & \multicolumn{1}{c|}{\textbf{U-Net}} & \multicolumn{1}{c|}{\textbf{CUnet}} & \multicolumn{1}{c|}{\textbf{UNETR}} & \multicolumn{1}{c|}{\textbf{Swin UNETR}} & \multicolumn{1}{c|}{\textbf{MSUneTr}} & \multicolumn{1}{c|}{\textbf{DeceptiConv}} & \multicolumn{1}{c}{\textbf{SwinConvNet}}  \\
 \specialrule{.1em}{.05em}{.05em}
 U-Net & $-$ & $0.8608 \pm 0.0044$ & $0.7796 \pm 0.0127$ & $0.8233 \pm 0.0068$ & $0.8300 \pm 0.0084$ & $0.8519 \pm 0.0024$ & $0.8511 \pm 0.0047$  \\
 CUnet & $0.8608 \pm 0.0044$ & $-$ & $0.7889 \pm 0.0131$ & $0.8381 \pm 0.0062$ & $0.8340 \pm 0.0062$ & $0.8609 \pm 0.0034$ & $0.8596 \pm 0.0032$    \\
 UNETR & $0.7796 \pm 0.0127$ & $0.7889 \pm 0.0131$ & $-$ & $0.8052 \pm 0.0118$ & $0.7847 \pm 0.0118$ & $0.7935 \pm 0.0136$ & $0.7963 \pm 0.0132$  \\
 Swin UNETR & $0.8233 \pm 0.0068$ & $0.8381 \pm 0.0062$ & $0.8052 \pm 0.0118$ & $-$ & $0.8206 \pm 0.0087$ & $0.8407 \pm 0.0049$ & $0.8447 \pm 0.0054$  \\
 MSUneTr & $0.8300 \pm 0.0084$ & $0.8340 \pm 0.0062$ & $0.7847 \pm 0.0118$ & $0.8206 \pm 0.0087$ & $-$ & $0.8290 \pm 0.0063$ & $0.8341 \pm 0.0056$  \\
 DeceptiConv & $0.8519 \pm 0.0024$ & $0.8609 \pm 0.0034$ & $0.7935 \pm 0.0136$ & $0.8407 \pm 0.0049$ & $0.8290 \pm 0.0063$ & $-$ & $0.8667 \pm 0.0034$  \\
 SwinConvNet & $0.8511 \pm 0.0047$ & $0.8596 \pm 0.0032$ & $0.7963 \pm 0.0132$ & $0.8447 \pm 0.0054$ & $0.8341 \pm 0.0056$ & $0.8667 \pm 0.0034$ & $-$  \\
\specialrule{.2em}{.1em}{.1em} 
\end{tabular}
}
\label{table:table3}
\end{table}



\begin{figure}[H]
\centering
\includegraphics[width=17cm,height=21cm]{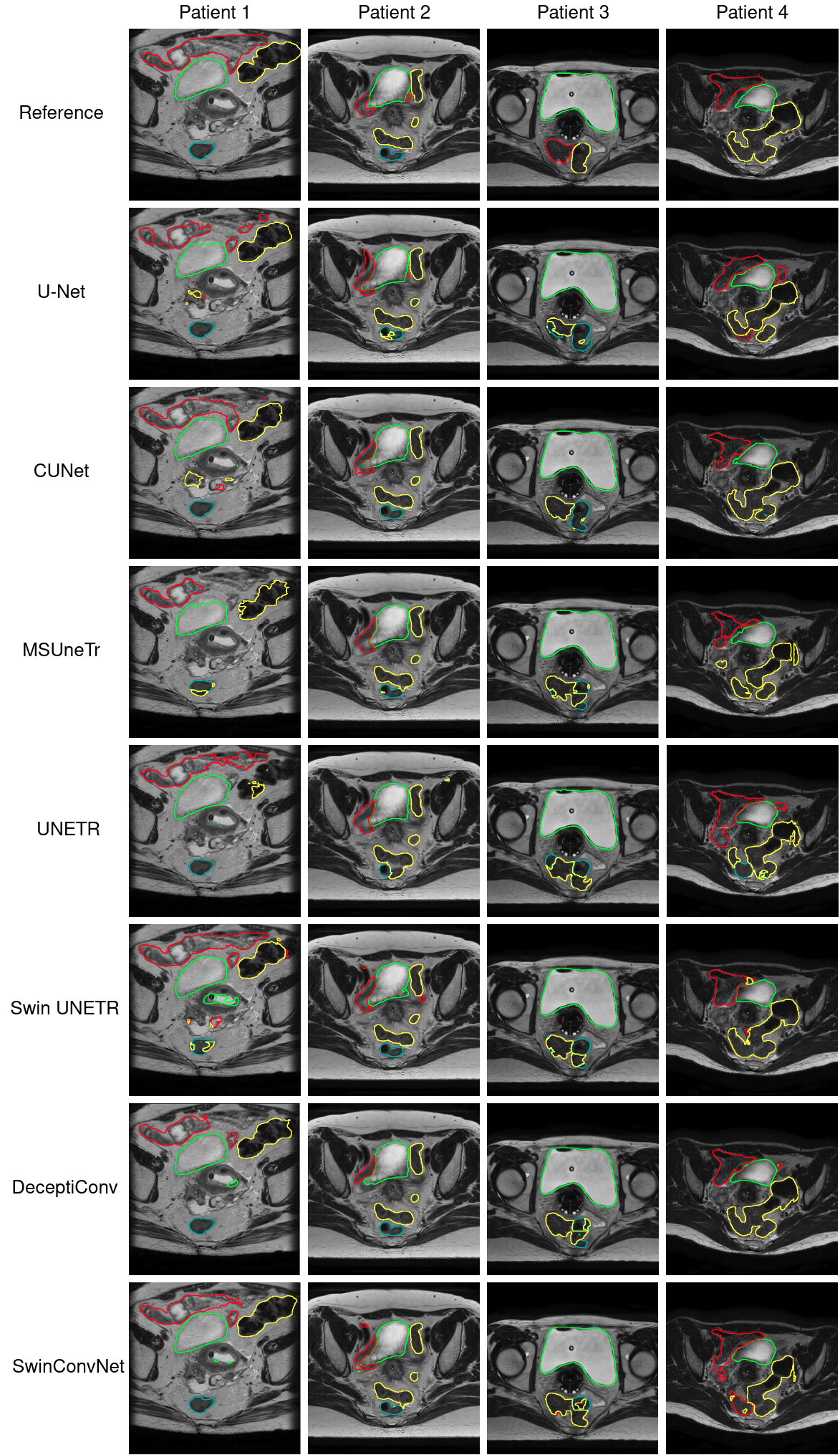}
\vspace{0.5cm}
\caption{\footnotesize{Predictions of models on the Cervical Cancer Dataset. {\color{mygreen} \textbf{Green}}: bladder, {\color{Red} \textbf{Red}}: bowel, {\color{mycyan} \textbf{Cyan}}: rectum, {\color{myyellow} \textbf{Yellow}}: sigmoid.}}
\label{fig:examples3}
\end{figure}

\begin{figure}[H]
\centering
\includegraphics[width=17cm,height=17cm]{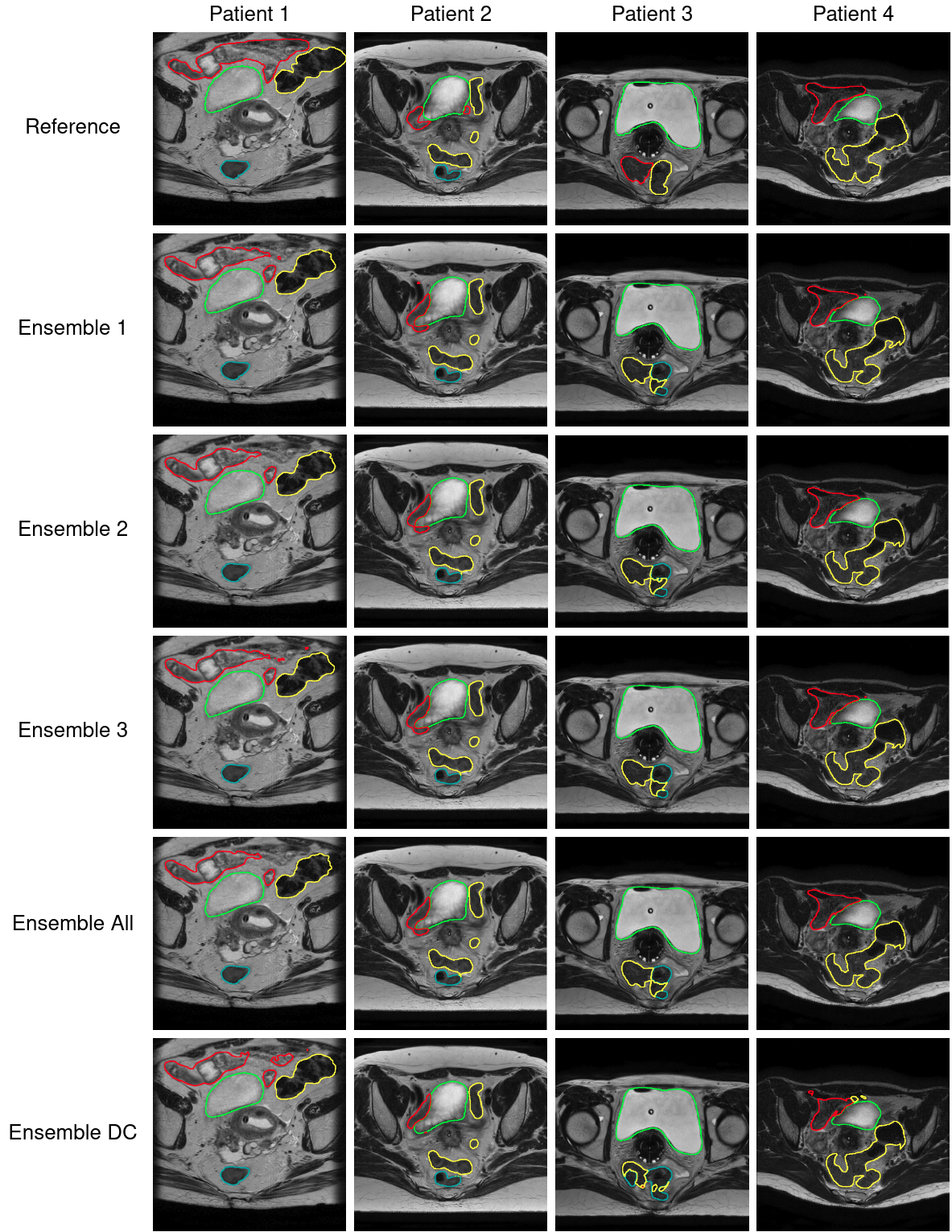}
\vspace{0.5cm}
\caption{\footnotesize{Predictions of ensembles of models on the Cervical Cancer Dataset. {\color{mygreen} \textbf{Green}}: bladder, {\color{Red} \textbf{Red}}: bowel, {\color{mycyan} \textbf{Cyan}}: rectum, {\color{myyellow} \textbf{Yellow}}: sigmoid.}}
\label{fig:examples4}
\end{figure}

For visualization purposes, we calculate the average Dice Coefficient of every model, for every round in their 5-fold experiment, resulting in 5 scores per model. Then we average all these values for every fold, resulting in 5 final values. We sort them, and keep the fold that corresponds to the median of these values, which in our case was the $2^{nd}$ fold. We randomly selected 4 patients from the validation set of this fold, from which we then selected 1 slice from each patient based on the reference that contains a lot of information. The results are presented in \autoref{fig:examples3} and \autoref{fig:examples4}.

Finally, for the statistical tests we collect all the average Dice scores across OARs for every patient, from all the folds in the 5-fold experiments. From the first test (DeceptiConv and Ensemble 1), we get a $p$-value of $\approx 1.5 \cdot10^{-18}$, while for the second test (Decepticonv and SwinConvNet) we get a $p$-value of $\approx 0.36$, indicating that we cannot really say that the two best performing models have a significant difference, but we can say with enough confidence that there is a difference between the best performing model and the best ensemble.

\section{Conclusions \& Discussion}
\label{sec:conclusions}

Assuming that we do not have datasets available for pretraining models, in this study, we followed the standard approach of manually tuning network architectures for the task of segmentation of OARs in MRI images of cervical cancer patients. Our goal was to investigate the differences in performance between various popular and novel architectures, and to find out if there is a singular potential architecture we can use to achieve the best results.

We introduced two new architerctures, the MSUneTr, and the DeceptiConv. The MSUneTr is a novel approach of using self-attention mechanisms to encode images for segmentation tasks. The main advantage is that with this architecture we can learn the attention that needs to be assigned between all patches of an image for a segmentation task, while using very small patch sizes and reasonably big enough image size. Another advantage is that we can parallelize the computations of the encoder, e.g., using one GPU per encoder layer, and thus achieve greater speed. Finally, the DeceptiConv is an interesting approach of combining Transformers with convolutions, as in this network's encoder both Transformers and convolutions follow their own path, and so they do not affect each other, which might be desired in some cases (e.g., if on a specific task Transformers do not perform well, then the network can omit the respective layers and focus only on convolutions, while in architectures like the SwinConvNet, both features are mixed and passed to every layer). 

Transformer-based models perform a little worse than CNN-based models on the Cervical Cancer Dataset, but this is not the case with the Brain Tumour Dataset, where all models achieve the same performance (except the UNETR model) (see \hyperref[sec:appendix]{appendix}). The Cervical Cancer Dataset is rather small, so this could be one reason why Transformers are not performing that good on this dataset. However, both the MSUneTr and the Swin UNETR are performing much better than the UNETR model, both in terms of performance metrics and visual results. 

We observed that for both the Cervical Cancer Dataset and the Brain Tumour Dataset that all models perform approximately equal according to quantitative performance metrics (except the UNETR in these specific tasks), but the combination of them obtained by ensembling does seem to further enhance the performance. The ensembles of different architectures can achieve slightly better performance than the ensemble of versions of the same architecture, which is an interesting finding. This could mean that eventually, in order to achieve the best possible performance, we might need to train different architectures, but also multiple times, and then create an ensemble from multiple versions of multiple architectures.

The visual results of the predictions of the models, and the Dice scores between their predictions (most of them achieving $> 0.82$ Dice score) also indicate a high similarity in performance, but at the same time every model makes different mistakes. We can observe a high confusion of the models when predicting the sigmoid, bowel, and rectum, for the reasons we mentioned earlier. In order to mitigate this problem, it might be wise to follow a different approach than trying to segment the OARs separately (e.g., segment these three organs as one organ and then separate them). Nonetheless, to eventually decide which model is better in practice and whether its results are clinically acceptable, we have to conduct validation studies with doctors, something we are planning to do in the near future.

\section*{ACKNOWLEDGMENTS}
\label{sec:acknowledgements}

The research is part of the research programme Open Technology Programme with project number 18373, which is financed by the Dutch Research Council (NWO), Elekta, and ORTEC LogiqCare. This work was carried out (in part) on the Dutch national e-infastructure with the support of SURF Cooperative.

\bibliography{spie_paper_bib} 

\begin{thebibliography}{10}

\bibitem{obsvar1}
Mukesh, M., Benson, R., Jena, R., Hoole, A., Roques, T., Scrase, C., Martin,
  C., Whitfield, G., Gemmill, J., and Jefferies, S., ``Interobserver variation
  in clinical target volume and organs at risk segmentation in
  post-parotidectomy radiotherapy: can segmentation protocols help?,'' {\em The
  British Journal of Radiology}~{\bf 85}(1016),  e530--e536 (2012).

\bibitem{obsvar2}
Vinod, S.~K., Jameson, M.~G., Min, M., and Holloway, L.~C., ``Uncertainties in
  volume delineation in radiation oncology: a systematic review and
  recommendations for future studies,'' {\em Radiotherapy and Oncology}~{\bf
  121}(2),  169--179 (2016).

\bibitem{obsvar3}
Saarnak, A.~E., Boersma, M., van Bunningen, B.~N., Wolterink, R., and
  Steggerda, M.~J., ``Inter-observer variation in delineation of bladder and
  rectum contours for brachytherapy of cervical cancer,'' {\em Radiotherapy and
  Oncology}~{\bf 56}(1),  37--42 (2000).

\bibitem{obsvar4}
Sharp, G., Fritscher, K.~D., Pekar, V., Peroni, M., Shusharina, N.,
  Veeraraghavan, H., and Yang, J., ``Vision 20/20: perspectives on automated
  image segmentation for radiotherapy,'' {\em Medical Physics}~{\bf 41}(5),
  050902 (2014).

\bibitem{unet}
Ronneberger, O., Fischer, P., and Brox, T., ``{U-Net}: Convolutional networks
  for biomedical image segmentation,'' in [{\em International Conference on
  Medical Image Computing and Computer-Assisted
  Intervention}{\nolinebreak\hspace{0.1em}]},   234--241, Springer (2015).

\bibitem{vnet}
Milletari, F., Navab, N., and Ahmadi, S.-A., ``{V-Net}: Fully convolutional
  neural networks for volumetric medical image segmentation,'' in [{\em 2016
  fourth International Conference on 3D Vision
  (3DV)}{\nolinebreak\hspace{0.1em}]},   565--571, IEEE (2016).

\bibitem{nnunet}
Isensee, F., Jaeger, P.~F., Kohl, S.~A., Petersen, J., and Maier-Hein, K.~H.,
  ``{nnU-Net}: a self-configuring method for deep learning-based biomedical
  image segmentation,'' {\em Nature Methods}~{\bf 18}(2),  203--211 (2021).

\bibitem{swinunetr2}
Tang, Y., Yang, D., Li, W., Roth, H.~R., Landman, B., Xu, D., Nath, V., and
  Hatamizadeh, A., ``Self-supervised pre-training of {Swin Transformers} for
  {3D} medical image analysis,'' in [{\em Proceedings of the IEEE/CVF
  Conference on Computer Vision and Pattern
  Recognition}{\nolinebreak\hspace{0.1em}]},   20730--20740 (2022).

\bibitem{vit}
Dosovitskiy, A., Beyer, L., Kolesnikov, A., Weissenborn, D., Zhai, X.,
  Unterthiner, T., Dehghani, M., Minderer, M., Heigold, G., Gelly, S.,
  Uszkoreit, J., and Houlsby, N., ``An image is worth 16x16 words: Transformers
  for image recognition at scale,'' in [{\em International Conference on
  Learning Representations}{\nolinebreak\hspace{0.1em}]},  (2021).

\bibitem{swintransformer}
Liu, Z., Lin, Y., Cao, Y., Hu, H., Wei, Y., Zhang, Z., Lin, S., and Guo, B.,
  ``Swin {Transformer}: Hierarchical vision {Transformer} using shifted
  windows,'' in [{\em Proceedings of the IEEE/CVF International Conference on
  Computer Vision}{\nolinebreak\hspace{0.1em}]},   10012--10022 (2021).

\bibitem{coatnet}
Dai, Z., Liu, H., Le, Q.~V., and Tan, M., ``{CoAtNet}: Marrying convolution and
  attention for all data sizes,'' {\em Advances in Neural Information
  Processing Systems}~{\bf 34},  3965--3977 (2021).

\bibitem{cvt}
Wu, H., Xiao, B., Codella, N., Liu, M., Dai, X., Yuan, L., and Zhang, L.,
  ``{CvT}: Introducing convolutions to vision {Transformers},'' in [{\em
  Proceedings of the IEEE/CVF International Conference on Computer
  Vision}{\nolinebreak\hspace{0.1em}]},   22--31 (2021).

\bibitem{ensemble}
Ganaie, M.~A., Hu, M., Malik, A., Tanveer, M., and Suganthan, P., ``Ensemble
  deep learning: A review,'' {\em Engineering Applications of Artificial
  Intelligence}~{\bf 115},  105151 (2022).

\bibitem{residuals}
He, K., Zhang, X., Ren, S., and Sun, J., ``Deep residual learning for image
  recognition,'' in [{\em Proceedings of the IEEE Conference on Computer Vision
  and Pattern Recognition}{\nolinebreak\hspace{0.1em}]},   770--778 (2016).

\bibitem{unetplus}
Zhou, Z., Rahman~Siddiquee, M.~M., Tajbakhsh, N., and Liang, J., ``Unet++: A
  nested {U-Net} architecture for medical image segmentation,'' in [{\em Deep
  Learning in Medical Image Analysis and Multimodal Learning for Clinical
  Decision Support}{\nolinebreak\hspace{0.1em}]},   3--11, Springer (2018).

\bibitem{deeplab}
Chen, L.-C., Papandreou, G., Kokkinos, I., Murphy, K., and Yuille, A.~L.,
  ``{DeepLab}: Semantic image segmentation with deep convolutional nets, atrous
  convolution, and fully connected {CRFs},'' {\em IEEE Transactions on Pattern
  Analysis and Machine Intelligence}~{\bf 40}(4),  834--848 (2017).

\bibitem{transformer}
Vaswani, A., Shazeer, N., Parmar, N., Uszkoreit, J., Jones, L., Gomez, A.~N.,
  Kaiser, L., and Polosukhin, I., ``Attention is all you need,'' {\em Advances
  in Neural Information Processing Systems}~{\bf 30} (2017).

\bibitem{unetr}
Hatamizadeh, A., Tang, Y., Nath, V., Yang, D., Myronenko, A., Landman, B.,
  Roth, H.~R., and Xu, D., ``{UNETR}: Transformers for {3D} medical image
  segmentation,'' in [{\em Proceedings of the IEEE/CVF Winter Conference on
  Applications of Computer Vision}{\nolinebreak\hspace{0.1em}]},   574--584
  (2022).

\bibitem{swinunetr}
Hatamizadeh, A., Nath, V., Tang, Y., Yang, D., Roth, H.~R., and Xu, D., ``Swin
  {UNETR}: {Swin} {Transformers} for semantic segmentation of brain tumors in
  {MRI} images,'' in [{\em International MICCAI Brainlesion
  Workshop}{\nolinebreak\hspace{0.1em}]},   272--284, Springer (2022).

\bibitem{msd}
Antonelli, M., Reinke, A., Bakas, S., Farahani, K., Kopp-Schneider, A.,
  Landman, B.~A., Litjens, G., Menze, B., Ronneberger, O., Summers, R.~M.,
  et~al., ``{The Medical Segmentation Decathlon},'' {\em Nature
  Communications}~{\bf 13}(1),  1--13 (2022).

\bibitem{transunet}
Chen, J., Lu, Y., Yu, Q., Luo, X., Adeli, E., Wang, Y., Lu, L., Yuille, A.~L.,
  and Zhou, Y., ``{TransUNet}: Transformers make strong encoders for medical
  image segmentation,'' {\em arXiv preprint arXiv:2102.04306}  (2021).

\bibitem{transfuse}
Zhang, Y., Liu, H., and Hu, Q., ``{TransFuse}: Fusing {Transformers} and {CNNs}
  for medical image segmentation,'' in [{\em International Conference on
  Medical Image Computing and Computer-Assisted
  Intervention}{\nolinebreak\hspace{0.1em}]},   14--24, Springer (2021).

\bibitem{cervix1}
Mohammadi, R., Shokatian, I., Salehi, M., Arabi, H., Shiri, I., and Zaidi, H.,
  ``Deep learning-based auto-segmentation of organs at risk in high-dose rate
  brachytherapy of cervical cancer,'' {\em Radiotherapy and Oncology}~{\bf
  159},  231--240 (2021).

\bibitem{cervix2}
Zhang, D., Yang, Z., Jiang, S., Zhou, Z., Meng, M., and Wang, W., ``Automatic
  segmentation and applicator reconstruction for {CT}-based brachytherapy of
  cervical cancer using {3D} convolutional neural networks,'' {\em Journal of
  Applied Clinical Medical Physics}~{\bf 21}(10),  158--169 (2020).

\bibitem{cervix3}
Yoganathan, S., Paul, S.~N., Paloor, S., Torfeh, T., Chandramouli, S.~H.,
  Hammoud, R., and Al-Hammadi, N., ``Automatic segmentation of magnetic
  resonance images for high-dose-rate cervical cancer brachytherapy using deep
  learning,'' {\em Medical Physics}~{\bf 49}(3),  1571--1584 (2022).

\bibitem{cervix4}
Li, Z., Zhu, Q., Zhang, L., Yang, X., Li, Z., and Fu, J., ``A deep
  learning-based self-adapting ensemble method for segmentation in
  gynecological brachytherapy,'' {\em Radiation Oncology}~{\bf 17}(1),  1--10
  (2022).

\bibitem{performer}
Choromanski, K., Likhosherstov, V., Dohan, D., Song, X., Gane, A., Sarlos, T.,
  Hawkins, P., Davis, J., Mohiuddin, A., Kaiser, L., Belanger, D., Colwell, L.,
  and Weller, A., ``Rethinking attention with performers,'' in [{\em
  International Conference on Learning
  Representations}{\nolinebreak\hspace{0.1em}]},  (2021).

\bibitem{se}
Hu, J., Shen, L., and Sun, G., ``Squeeze-and-excitation networks,'' in [{\em
  Proceedings of the IEEE Conference on Computer Vision and Pattern
  Recognition}{\nolinebreak\hspace{0.1em}]},   7132--7141 (2018).

\bibitem{transconver}
Liang, J., Yang, C., Zeng, M., and Wang, X., ``{TransConver}: Transformer and
  convolution parallel network for developing automatic brain tumor
  segmentation in {MRI} images,'' {\em Quantitative Imaging in Medicine and
  Surgery}~{\bf 12}(4),  2397 (2022).

\end{thebibliography}
\bibliographystyle{spiebib} 

\newpage

\section*{APPENDIX - Experiments on the Brain Tumour Dataset}
\label{sec:appendix}

\begin{figure}[H]
\centering
\includegraphics[width=17cm,height=6cm]{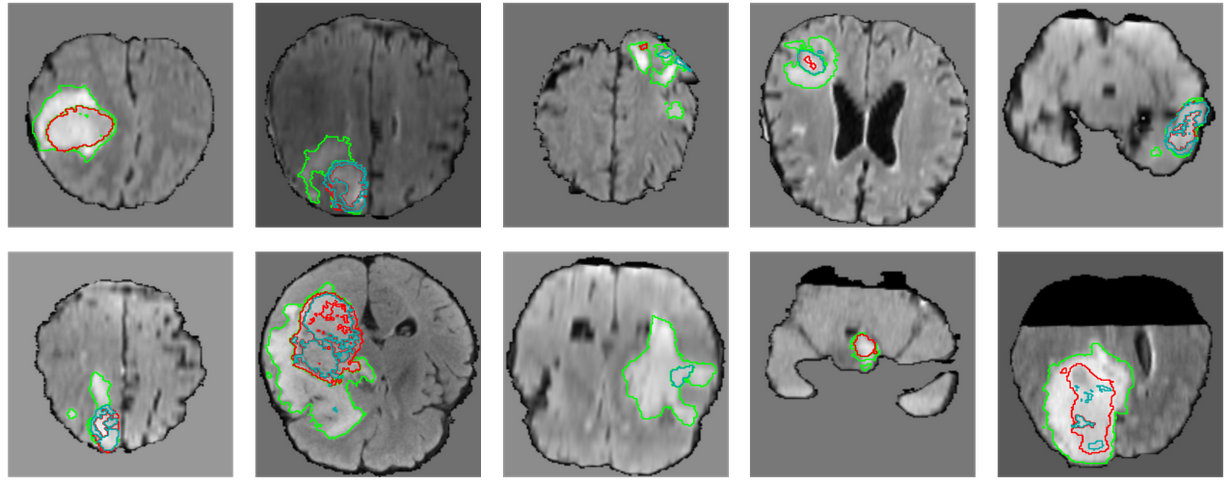}
\vspace{0.5cm}
\caption{\footnotesize{Slices of images from the Brain Tumour Dataset. {\color{mygreen} \textbf{Green}}: edema, {\color{Red} \textbf{Red}}: non-enhancing tumour, {\color{mycyan} \textbf{Cyan}}: enhancing tumour.}}
\label{fig:examples2}
\end{figure}

For the purpose of, validating the results in other tasks, and also increasing the transparency of our work, we have repeated the same study for the Brain Tumour Dataset of the Medical Segmentation Decathlon \cite{msd}, and observed similar findings. The data consist of 750 multiparametric Magnetic Resonance Images (mp-MRI) from patients diagnosed with either glioblastoma or lower-grade glioma, from which we used only the training set (484 scans). The sequences used were native T1-weighted (T1), post-Gadolinium (Gd) contrast T1-weighted (T1-Gd), native T2-weighted (T2), and T2 Fluid-Attenuated Inversion Recovery (FLAIR). The corresponding target regions of interest are the three tumour sub-regions, namely edema, enhancing tumour, and non-enhancing tumour. The data was acquired from 19 different institutions and contained a subset of the data used in the 2016 and 2017 Brain Tumour Segmentation (BraTS) challenges. The scans consist of a total of 66517 2D slices, which were used as the training dataset (\autoref{fig:examples2}).

Most of the training details are the same as those used with the Cervical Cancer Dataset, so we will mention only the differences. For the Brain Tumour Dataset we used a crop size of $128 \times 128$. Preprocessing and augmentation were the same as on the Cervical Cancer Dataset, with the only difference being that we also used flips. Fifty epochs were used to train the models on the Brain Tumour Dataset. 

Technically, the SwinConvNet achieved the best average performance across classes in this case according to the metrics. Therefore, we trained the SwinConvNet 5 additional times to create an ensemble from versions of this model, which we call Ensemble SC. Regarding the ensembling of different architectures, the combination that achieved the best results is referred to as Ensemble 1 (DeceptiConv, SwinConvNet, Swin UNETR, and MSUneTr), followed by Ensemble 2 (DeceptiConv, CUnet, Swin UNETR, MSUneTr, UNETR, and U-Net), Ensemble 3 (DeceptiConv, SwinConvNet, CUnet, Swin UNETR, and MSUneTr), and finally Ensemble All (all 7 models). 

\begin{figure}[H]
\centering
\includegraphics[width=17cm,height=7cm]{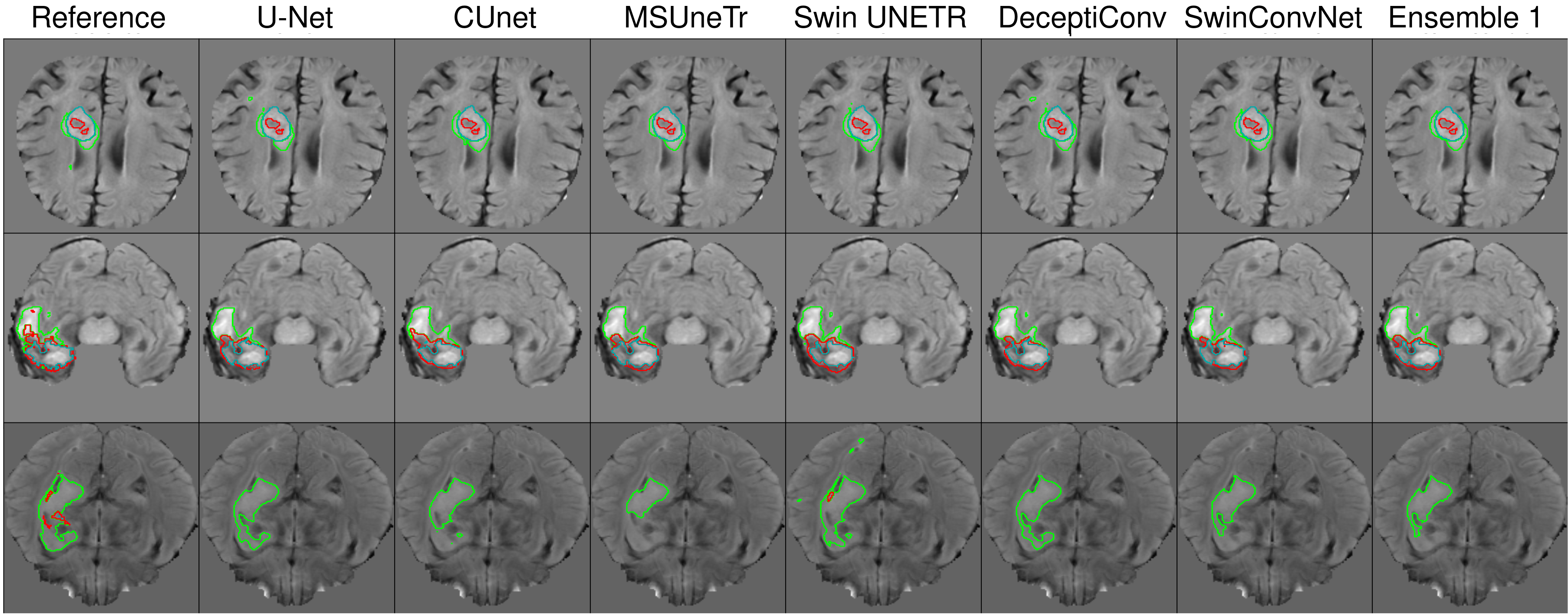}
\vspace{0.5cm}
\caption{\footnotesize{Predictions of models on the Brain Tumour Dataset. {\color{mygreen} \textbf{Green}}: edema, {\color{Red} \textbf{Red}}: non-enhancing tumour, {\color{mycyan} \textbf{Cyan}}: enhancing tumour.}}
\label{fig:examples5}
\end{figure}

The results from the 5-fold experiments and ensembling are presented in \autoref{table:table4}. For comparison with a state-of-the-art model, in \autoref{table:table4} we provide the results from the 2D network of \citenum{nnunet}. For visualization, we followed the same approach as on the Cervical Cancer Dataset to find the fold with the median performance of the models. Then we randomly selected 3 patients from the validation set of this fold, from which we selected 1 slice based on the reference with enough information. In \autoref{fig:examples5} we illustrate the predictions of the models on these 3 patients, where again we observe a high overlap between them, but at the same time every model makes different mistakes. We also calculated the Dice score between predictions of pairs of models (\autoref{table:table5}), and again observed a very high similarity (in most cases except those involving the UNETR model, the models achieve more than 0.84 Dice score).

At last, we conduct a Wilcoxon paired test for the best ensemble and the best performing model (the SwinConvNet), and for the two best performing models (DiceptiConv and SwinConvNet). Results with $p$-value $< 0.05$ were considered to be significant. Again, for these tests we collect all the Dice scores, across all OARs, for all the patients, from all the folds in the 5-fold experiments. From the first test, we get a $p$-value of $\approx 6.8 \cdot10^{-32}$, while for the second test we get a $p$-value of $\approx 0.93$, indicating again that we cannot really say that the two best performing models have a significant difference, but we can say with enough confidence that there is a difference between the best performing model and the best ensemble.

\begin{table}[H]
\captionsetup{font=footnotesize, width=0.85\textwidth}
\centering
\caption{Average Dice Coefficient and $95^{th}$ percentile of Hausdorff Distance (HD) in mm, on the validation sets, from models trained with 5-fold training, and the best performing ensembles on the Brain Tumour Dataset.}
\resizebox{0.85\textwidth}{!}{%
\begin{tabular}{*{1}{>{\bfseries}l} *{8}{|>{\bfseries}c} }
 \specialrule{.2em}{.1em}{.1em} 
 \multicolumn{1}{c|}{Model} & \multicolumn{2}{c|}{edema} & \multicolumn{2}{c|}{non-enh. tumour} 
 & \multicolumn{2}{c|}{enhancing tumour} & \multicolumn{2}{c}{Avg} \\
 \specialrule{.1em}{.05em}{.05em}
  & Dice & HD & Dice & HD & Dice & HD & Dice & HD \\
 \specialrule{.1em}{.05em}{.05em}
 nnU-Net 2D (\citenum{nnunet}) & $0.7957$ & $-$ & $0.5985$ & $-$ & $0.7825$ & $-$ & $0.7256$ & $-$  \\
 \specialrule{.1em}{.05em}{.05em}
 U-Net & $0.7857$ & $8.4981$ & $0.6020$ & $8.4115$ & $0.7832$ & $6.0893$ & $0.7237$ & $7.7164$  \\
 CUnet & $0.7870$ & $9.3802$ & $0.5977$ & $8.7814$ & $0.7810$ & $6.0892$ & $0.7219$ & $8.1487$  \\
 UNETR & $0.7738$ & $16.3932$ & $0.5725$ & $10.6124$ & $0.7637$ & $8.1568$ & $0.7033$ & $11.8963$ \\
 Swin UNETR  & $0.7879$ & $14.3805$ & $0.6025$ & $9.5052$ & $0.7829$ & $6.5107$ & $0.7244$ & $10.3218$ \\
 MSUneTr & $0.7858$ & $9.6683$ & $0.6030$ & $8.9293$ & $0.7833$ & $6.3999$ & $0.7240$ & $8.4513$ \\
 DeceptiConv & $0.7911$ & $9.6990$ & $0.6041$ & $8.0622$ & $0.7803$ & $5.8192$ & $0.7252$ & $7.9448$ \\
 SwinConvNet & $0.7916$ & $9.3137$ & $0.6029$ & $8.3294$ & $0.7853$ & $6.0829$ & $0.7266$ & $8.0087$ \\
 \specialrule{.1em}{.05em}{.05em}
 Ensemble SC & $0.7970$ & $8.9494$ & $0.6106$ & $8.2966$ & $0.7925$ & $6.1606$ & $0.7333$ & $7.8022$  \\
 \specialrule{.1em}{.05em}{.05em}
 Ensemble 1 & $0.7995$ & $9.0480$ & $0.6163$ & $7.9728$ & $0.7938$ & $5.4113$ & $0.7366$ & $7.4774$  \\
 Ensemble 2 & $0.8001$ & $8.0927$ & $0.6173$ & $7.7527$ & $0.7917$ & $5.4700$ & $0.7364$ & $7.1051$  \\
 Ensemble 3 & $0.8001$ & $8.6563$ & $0.6166$ & $7.8053$ & $0.7925$ & $5.4444$ & $0.7364$ & $7.3020$  \\
 Ensemble All & $0.8007$ & $8.3950$ & $0.6174$ & $7.6468$ & $0.7908$ & $5.5654$ & $0.7363$ & $7.2024$  \\
\specialrule{.2em}{.1em}{.1em} 
\end{tabular}
}
\label{table:table4}
\end{table}



\begin{table}[H]
\captionsetup{font=footnotesize, width=\textwidth}
\centering
\caption{Average Dice Coefficient on the validation sets, between predictions of models trained with 5-fold training on the Brain Tumour Dataset.}

\resizebox{\textwidth}{!}{%
\begin{tabular}{*{1}{>{\bfseries}l} *{7}{|>{\bfseries}c} }
 \specialrule{.2em}{.1em}{.1em} 
 \multicolumn{1}{c|}{} & \multicolumn{1}{c|}{\textbf{U-Net}} & \multicolumn{1}{c|}{\textbf{CUnet}} & \multicolumn{1}{c|}{\textbf{UNETR}} & \multicolumn{1}{c|}{\textbf{Swin UNETR}} & \multicolumn{1}{c|}{\textbf{MSUneTr}} & \multicolumn{1}{c|}{\textbf{DeceptiConv}} & \multicolumn{1}{c}{\textbf{SwinConvNet}} \\
 \specialrule{.1em}{.05em}{.05em}
 U-Net & $-$ & $0.8513$ & $0.7997$ & $0.8337$ & $0.8432$ & $0.8434$ & $0.8450$ \\
 CUnet & $0.8513$ & $-$ & $0.8008$ & $0.8380$ & $0.8483$ & $0.8502$ & $0.8495$ \\
 UNETR & $0.7997$ & $0.8008$ & $-$ & $0.8168$ & $0.8093$ & $0.8134$ & $0.8115$ \\
 Swin UNETR & $0.8337$ & $0.8380$ & $0.8168$ & $-$ & $0.8462$ & $0.8459$ & $0.8522$ \\
 MSUneTr & $0.8432$ & $0.8483$ & $0.8093$ & $0.8462$ & $-$ & $0.8581$ & $0.8587$ \\
 DeceptiConv & $0.8434$ & $0.8502$ & $0.8134$ & $0.8459$ & $0.8581$ & $-$ & $0.8595$ \\
 SwinConvNet & $0.8450$ & $0.8495$ & $0.8115$ & $0.8522$ & $0.8587$ & $0.8595$ & $-$ \\
\specialrule{.2em}{.1em}{.1em} 
\end{tabular}
}
\label{table:table5}
\end{table}

\end{document}